\documentclass[12]{article}
\usepackage{graphicx,float}
\usepackage{caption}
\usepackage{subcaption}
\usepackage{amsmath}
\usepackage{breqn}
\usepackage{multicol}
\usepackage{mathtools}
\usepackage{authblk}
\usepackage{float}
\usepackage[margin=0.8in]{geometry}
\usepackage{cite}
\usepackage[square,numbers]{natbib}
\usepackage{sectsty}
\usepackage{indentfirst}
\usepackage[font={small}]{caption}
\usepackage{csquotes}

\sectionfont{\fontsize{10}{15}\selectfont\centering}
\subsectionfont{\fontsize{10}{0}\selectfont\centering}
\usepackage{titlesec}
\titleformat*{\subsubsection}{\normalfont\fontfamily{phv}\fontsize{14}{17}\selectfont}
\fontfamily{phv}
\AtBeginDocument{}
\usepackage[hidelinks,colorlinks=true,linkcolor=blue,citecolor=blue]{hyperref}
\hypersetup{
    hidelinks=true,
    colorlinks=true,
    linkcolor=blue,
    filecolor=magenta,      
    urlcolor=blue,
    }
\begin{document}
\title{\textbf{Detailed Analysis of Two Simplified Rydberg Dressed Atoms Confined in a Harmonic Trap} }
\author[1]{Leila Chia$^{2,}$}
\author[2]{Nabila Grar\thanks{corresponding author: \href{mailto:yourmail}{\color{blue}{nabila.grar@univ-bba.dz}} }}

%\author{Nabila Grar$^1$ and Leila Chia$^{2,}$}
%\ead{nabila.grar@bba-univ.dz}

\affil[1]{\small{Laboratory of Materials Physics, Radiation and Nanostructures (LPMRN),
Faculty of Sciences and Technology, Mohamed El Bachir El Ibrahimi
University, Bordj-Bou-Arreridj 34030, Algeria}}
\affil[2]{\small {Department of sciences of matter, Faculty of Sciences and Technology,
		El Bachir El Ibrahimi University, Bordj Bou Arreridj 34030, Algeria}}
\maketitle
\providecommand{\keywords}[1]{\textbf{\textit{Index terms---}} #1}	
	\begin{abstract} By using a step-like potential, it is possible to mimic the Rydberg short range part of the interaction between two atoms. It is easy in this case to establish an analytical solution of the Schr\"{o}dinger equation. In this contribution, we are analyzing in detail this simplified model by highlighting the major players in different interaction schemes (strengths and ranges), different dimensionalities and the impact on spatial correlation. We are able to achieve an improvement to this model by applying a perturbation treatment to the potential. The dynamical aspects related to a sudden change of the potential features are also investigated.
	\end{abstract}

	\section{Introduction}
	\label{section1} %label of introduction section
	Matter is a huge and intricate assembly of some \enquote{fundamental} constituents where seemingly the individuality of these constituents is lost. The understanding of the finite micro elements leading to the macroscopic structure is however of paramount importance for both nuclear and condensed matter physics. It aims not only to the comprehension of the constituent's structure but also to the elucidation of the correlations and the interplay among constituents.  From another side, experiments involving confined cold few particles are nowadays very accessible and are becoming a matter of routine. In fact  number of the systems parameters as the confinement potential and the particle-particle interaction features can be controlled on demand\cite{blum,abra}. This way, it is possible to verify experimentally the validity of a  number of quantum simplified models studied by the past and explore fundamental physics concepts. It is also true that more efforts are necessary in order to devise new ``toy`` models targeting a detailed comprehension of the features of the interaction at different levels of approximation as well as different dimensionalities (1D, 2D and 3D).\\ 
	\noindent
	Exact solution for the Schr\"{o}dinger equation established in the case of different and sometimes complicated potentials, can be found in the literature (see for example the Refs. \cite{turb,mora,gao1,gao2,gino}). Most of these solutions are given for the case of  single particle system. The situation becomes quite complicated when considering the case of two particles as a first step on the path towards the description of cold confined mesoscopic systems \cite{blai, pita, koci1}. The difficulty resides in the consideration of both confinement potential characteristics and realistic interaction potential. The hard core interaction is the most simplified interaction scheme  and in this case, it  is possible to achieve a quasi exact solution for the two particles system. A theoretical work encompassing the three dimensionalities and a delta-like interaction for a system of two particles was elaborated in the seminal work of Busch et al. \cite{busc, wei}. A quasi exact solution is hence established where the interaction is considered to be of contact nature (an s-wave  for bosons and p-wave  for fermions). In order to take into account a certain interaction range, a Gaussian-like potential can be considered and in this case also a quasi analytical solution can be achieved \cite{doga,muja}. These interaction models however ignore the long range nature of interaction for dipolar atoms or the  Rydberg dressed interaction behaving like $1/r^6$ and which can be very important either fundamentally or experimentally \cite{oldz,lim}. An analytical solution for this interaction is still to be found. Nevertheless a simplification of this interaction as a step function was proposed by  Ko{\'s}cik et { al.} \cite{koci2}. It is possible in this case to reach a quasi exact solutions in one and two dimensions and a study of the different features of the system was elaborated. This kind of  quasi solvable models are of extreme importance for advances in cold confined few particles systems. It can be considered as a set of models to be validated  experimentally  as well as an exact basis to construct the solution for few body systems exploiting different strategies (as variational, ab initio, interacting configurations...) \cite{blai,pita,koci1}.\\ 
	\noindent
	The aim of our present study is to elaborate a comparative analysis of the quasi exact solvable model of  Ko{\'s}cik et al. in the three dimensionalities and highlight the most important players for the  considered interaction. We are concerned unavoidably by the analysis of the spatial correlation as an important part of the information about the studied system. We propose an improvement to the cited model by applying a perturbation treatment to the potential. We are also addressing the most important results of the dynamical evolution of the system under a sudden change of the potential features and how this evolution is affecting the correlation.\\
	\noindent
	The paper is organized as follows. In the second section we are recalling the most important formula to be used in the following sections. In the third section we analyze the energy spectra for the relative part of the Schr\"{o}dinger equation solution for the system of two particles for different schemes  (strength and range) of the interaction. The aim is to single out the major players for different dimensionalities. In a fourth section the relative-radial spatial correlation is analyzed for the three dimensionalities. The interplay between the centrifugal repulsive effect, the range and the strength of the interaction is studied in detail. In the fifth section we are proposing an improvement to Ko{\'s}cik et al. model by exploiting a perturbation treatment of the potential. In the sixth section a dynamical study of the system is elaborated and the effects of the sudden change of the interaction parameters on the system correlation is investigated. The main results are summarized in the final conclusion.
	
	\section{Theoretical approach}
	\label{section2} % label of section 2
	The aim of the different models is to establish an analytical solution for the following Schr\"{o}dinger equation for a system of two identical spinless quantum particles, having a mass $m$ and trapped in an external potential:
	\begin{equation} 
	\left(  \sum^{2}_{i=1}\frac{-\hbar^{2}}{2m}\nabla^{2}_{i}+v_{ext}+ v \right)  \psi(\overrightarrow{r_1} ,\overrightarrow{r_2} )=E \psi(\overrightarrow{r_1} ,\overrightarrow{r_2} ), \label{eq:eq1}
	\end{equation}
	where, $ v_{ext}$ is the confining potential, $ v$ is the interaction potential depending on the particles separation and $\overrightarrow{r_i} $ is the vector position for each particle. To simplify the calculation, the two particles are considered to be structureless i.e point-like and the confining potential is considered to be harmonic. The anistropy considered in  the harmonic oscillator defines the constraint on the motion of the particles and consequently defines the dimensionality of the problem\cite{zinn,Isla,gor}. The same confining potential is imposed to both particles and the equation becomes:
	\begin{equation} 
	%\begin {split}
	\left( \left( \sum^{2}_{i=1}\frac{-\hbar^{2}}{2m}\nabla^{2}_{i}+ \frac{1}{2} m\omega^2 r_i^2 \right)+ v(\vert\overrightarrow{r_1}-\overrightarrow{r_2}\vert) \right) \psi(\overrightarrow{r_1} ,\overrightarrow{r_2} ) =  E \psi(\overrightarrow{r_1} ,\overrightarrow{r_2} ). \label{eq:eq2}
	% \end{split}
	\end{equation}
	For this quadratic potential, it is  possible to single out the center of mass  contribution to the motion from the relative one  and  we can write the equation as: 
	
	\begin{equation} 
	%\begin {split}
	\left( \frac{-\hbar^{2}}{2M}\nabla^{2}_{\overrightarrow{R}}+ \frac{1}{2} M\omega^2 R^2 +\frac{-\hbar^{2}}{2\mu}\nabla^{2}_{\overrightarrow{r}}+ \frac{1}{2} \mu\omega^2 r^2 + v(r) \right)   \psi(\overrightarrow{R},\overrightarrow{r}) =E\psi(\overrightarrow{R},\overrightarrow{r}), \label{eq:eq3}
	%\end{split} 
	\end{equation}
	where, $ M=2m$, $ \mu =m/2$ (the reduced mass), $R=\vert\overrightarrow{r_1}+\overrightarrow{r_2}\vert/2$  and $ r=\vert \overrightarrow{r_1}-\overrightarrow{r_2}\vert$. 
	The wave function can be written in a separable form  as:
	\begin{equation} 
	\psi(\overrightarrow{R},\overrightarrow{r})=\chi (\overrightarrow{R})\varphi(\overrightarrow{r}). \label{eq:eq4}
	\end{equation}
	Consequently we can separate the center of mass motion from the relative one  as:
	\begin{equation} 
	\left( \frac{-\hbar^{2}}{2M}\nabla^{2}_{\overrightarrow{R}}+ \frac{1}{2} M\omega^2 R^2 -E_c \right)  \chi(\overrightarrow{R})=0, \label{eq:eq5}
	\end{equation}
	\begin{equation} 
	\left(  \frac{-\hbar^{2}}{2\mu}\nabla^{2}_{\overrightarrow{r}}+ \frac{1}{2} \mu\omega^2 r^2 + v(r) -E_r \right)\varphi(\overrightarrow{r})=0, \label{eq:eq6}
	\end{equation}
	\noindent
	with $E=E_c+E_r$. 
	The first equation is just an equation for a harmonic oscillator with known solutions. The difficulty resides in finding a solution for the second equation where handling a realistic interaction can be quite  challenging. It is important to remind here that the symmetry of the total wave function is dependent only on the relative part of the wave function since the center of mass part is symmetric by construction.
	In one dimension (say  for example $x=x_1-x_2$) the relative equation reduces to :
	\begin{equation} 
	\left(  \frac{-d^2}{dx^2}+ \frac{1}{4} x^2 + v(x) -E_r \right)\phi(x)=0. \label{eq:eq7}
	\end{equation}
	Here, the equation is written such as the energy and the position are expressed in $\hbar \omega$ and  $\sqrt{\frac{\hbar}{m \omega}}$  units respectively.
	For two dimensions we convert to polar coordinates $\overrightarrow{r} \rightarrow (r,\varphi) $ and with writing the relative wave function as:
	
	\begin{equation} 
	\phi(r,\varphi)=\frac{f(r)}{\sqrt{r}}e^{\pm i l \varphi},\label{eq:eq8}
	\end{equation}
	the equation for the relative-radial motion becomes :
	\begin{equation} 
	\left(  \frac{-d^2}{dr^2}+ \frac{l^2-1/4}{r^2} +\frac{1}{4} r^2 + v(r) -E_r \right)f(r)=0, \label{eq:eq9}
	\end{equation}
	\noindent
	where, $ l$  is the angular momentum quantum number and it is expressed in  $\sqrt{\hbar m \omega}$ units. The second term in this equation represents the centrifugal potential.
	For three dimensions we use spherical coordinates $\overrightarrow{r}  \rightarrow (r,\theta,\varphi)$ and the relative wave function is written as:
	\begin{equation} 
	\phi(r,\theta,\varphi)=\frac{1}{r} f(r) y_l^m(\theta,\varphi), \label{eq:eq10}
	\end{equation}
	$y_l^m(\theta,\varphi)$ being the  spherical harmonics.
	\noindent
	The equation for the relative-radial part is then given as :
	\begin{equation} 
	\left(  \frac{-d^2}{dr^2}+ \frac{l(l+1)}{r^2} +\frac{1}{4} r^2 + v(r) -E_r \right)f(r)=0. \label{eq:eq11}
	\end{equation}
	Notice here that we can shift from equation (\ref{eq:eq9}) to (\ref{eq:eq11}) by operating the following change:
	\begin{equation} 
	l_{2D} \rightarrow  l_{3D}+1/2.\label{eq:eq12}
	\end{equation}
	This will signify that it is possible to find the solution for the 3D case by just solving the equation for the 2D case but with respecting the previous relation between the two angular quantum numbers \cite{koci2}. The wave function is dependent on a  quantum number $n$ in one 1D, on ($n$ ,$l$) for the relative part of the wave function in 2D and on ($n$,$l$,$m$) for 3D. The total wave function however is symmetric for even $n$ in 1D  and even $l$ for 2D and 3D; and is antisymmetric for odd $n$ in 1D  and odd $l$ for 2D and 3D. As known a symmetric total wavefunction  defines a bosonic state and conversely an antisymmetric total  wave function defines a fermionic one.\\
	\begin{figure}[h!]
		\centerline{\includegraphics*[height=2.5in,width=2.7in, keepaspectratio]{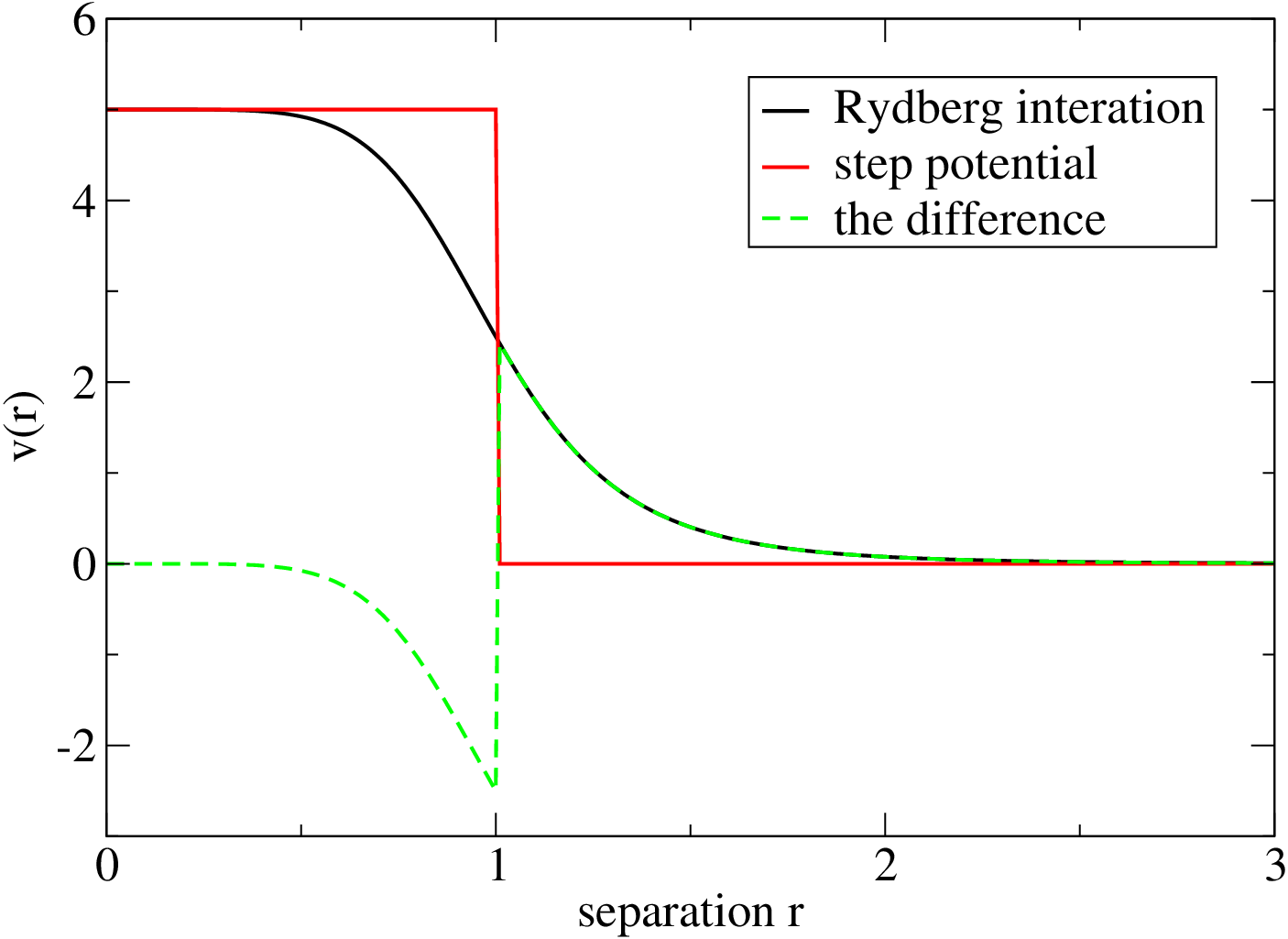} }
		\vspace*{8pt}
		\caption{Comparison between the realistic Rydberg interaction, the step function and the difference between these two potentials. $v_0$ is considered to be equal to 5 and the range is equal to one. The difference is considered as a perturbation (see sect.\ref {sec:5}).
		}
		\label{fig:fig1}
	\end{figure}
	
	\noindent
	Solving the equations (\ref{eq:eq7}) ,(\ref{eq:eq9}) and (\ref{eq:eq11}) will rely on the form considered for the interaction potential. As matter of fact, different forms of interaction are considered and studied in the literature. Starting from the hard core or contact-like potential defined as :

	\begin{equation}
	v(r)=\begin{dcases*}
	g_{hc} \delta (x-x_0) & $ ~for~ x \ge 0, $\\
	g_{hc} \delta(r-r_0)\frac{\partial}{\partial r}r & $~for~ 2~and~3~dimensions,$\\
	\end{dcases*}
	\label{eq:eq13} 
	\end{equation}
	
	\noindent
	where $g_{hc}$ is the strength of the hard core interaction. We notice that the delta function must be regularized in order to avoid singularities in two and three dimensions\cite{busc, wei}. Taking this into account, it is possible to solve quasi exactly the problem for the three dimensionalities.
	In order to introduce a certain finite range within the interaction, the considered potential can be a Gaussian shaped one given by: \cite{doga,muja}:
	\begin{equation}
	v(r) = \frac{g_g}{s^{2}}exp  \frac{\vert \overrightarrow{r_2}-\overrightarrow{r_1}\vert^{2} }{s^{2}},  
	\label{eq:eq14}  
	\end{equation}
	where, $g_g$ is the Gaussian interaction strength and $s$ is its range. It is possible in this case to achieve an approximate analytical formula that gives the energy spectra and allows the study of the effect of the finite range interaction on the system features. To take into account the long range nature of the interaction, otherwise the interaction between two non symmetric neutral charged distributions, we should consider excitations with multipolar proprieties. When atoms are excited to large principal numbers, these are known as  Rydberg states. The potential for this interaction can be approximated to the first order as composed of a short ranged part to which we add a van der waal long ranged interaction. This last one is the main contribution to the multipolar excitations. In this case the interaction potential can be given as \cite{koci2, dali}:
	\begin{equation} 
	v(r)=\frac{g}{1+\left( \frac{r}{R_c}\right)^6 }~,\label{eq:eq15}
	\end{equation}
	where, $g$ gives the strength and $R_c$ is the range of the potential respectively (see figure \ref{fig:fig1}). We will call this potential the Rydberg interaction in the following sections.
	It is not possible yet to find an exact solution to the equations (\ref{eq:eq7}), (\ref{eq:eq9}) and (\ref{eq:eq11}) with this realistic interaction along with the harmonic confinement, nonetheless a quasi exact solution is achieved for a potential defined as a step function \cite{koci2}. This approximation mimics the previous expression quite fairly for the short range part and then falls abruptly to zero. It is given as:

	\begin{equation}
	v(r)=\begin{dcases*}
	v_0  & $for~ r  \le a,$ \\
	0 & $for ~r>a,$\\
	\end{dcases*}
	\label{eq:eq16} 
	\end{equation}
	\noindent
	where we can relate $v_0$  and $a$ to the strength and the range ($g $ and $R_C$) respectively (figure \ref{fig:fig1}). This simplification is justified by the fact that the main contribution to the realistic potential comes from the flat part. In the case of this approximation, it is possible to establish a quasi exact solution by reducing the 1D equation (Eq. \ref{eq:eq7}) to a Weber form and to a Kummer form for the relative-radial equation in the case of 2D and 3D (Eqs. \ref{eq:eq9} and \ref{eq:eq11}). The solution is expressed as function of the confluent hypergeometric function of the first kind in the  region $[0,a]$ and  as function of the Tricomi function elsewhere \cite{koci2, math, abra2,digi}. In order to guarantee a physical behavior of the whole solution, a condition for the continuity of the two functions and their derivatives is imposed at $r=a$  and hence this gives rise to transcendental equations. The solution of these equations  leads to the quantification of the energy which allows to retrieve the energy spectrum with different combinations of strength $v_0$  and the range $a$.

	\section{Energy spectrum and the dimensionality impact  }
	\label{section3} % label of section 3
	
\begin{figure}[h!]
		\centering
		\begin{subfigure}[t]{.5\textwidth}
			\centering
			\includegraphics*[width=.8\linewidth]{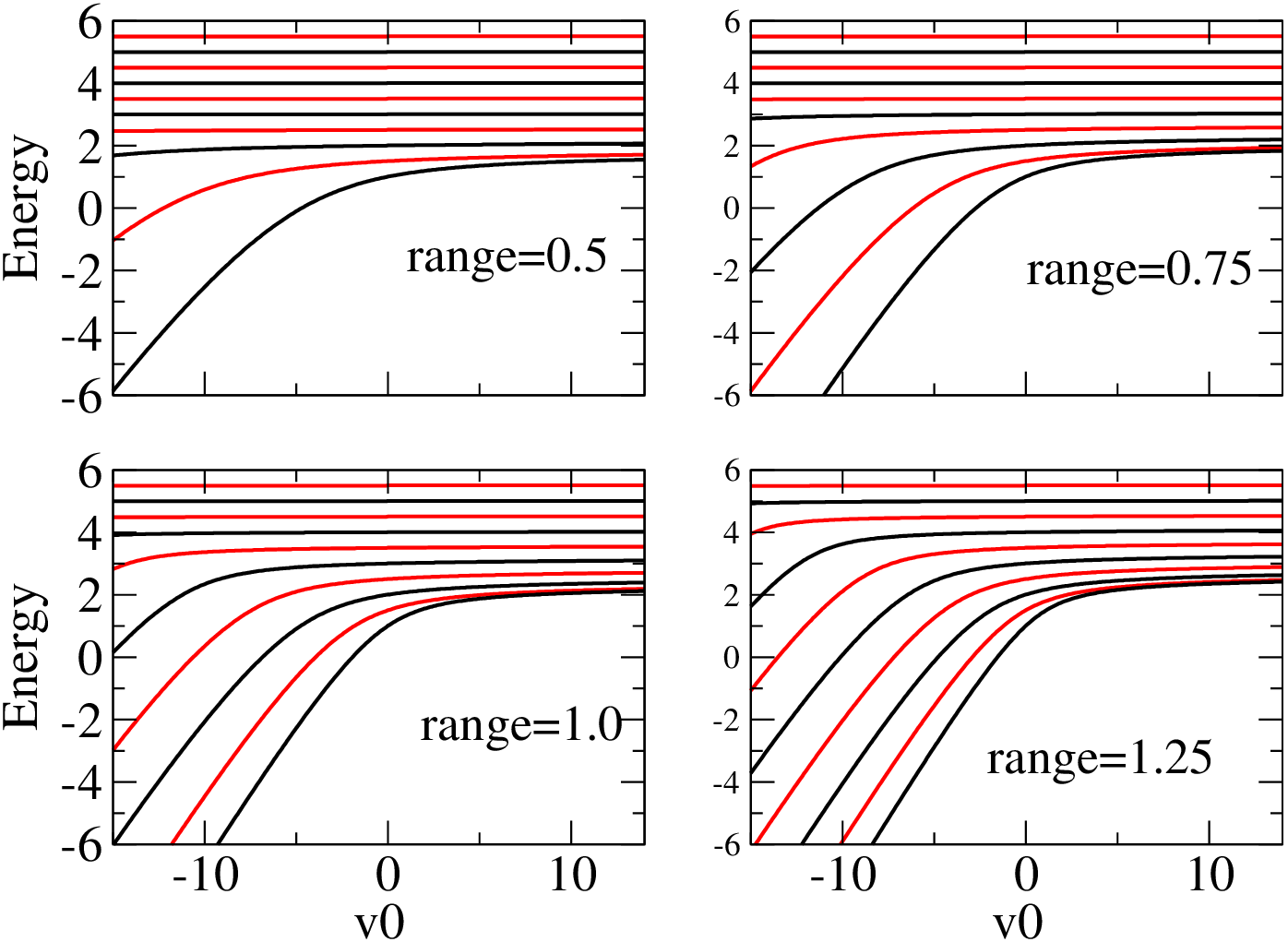}
			\caption{}
			\label{fig:sub10}
		\end{subfigure}%
		\begin{subfigure}[t]{.5\textwidth}
			\centering
			\includegraphics*[width=.8\linewidth]{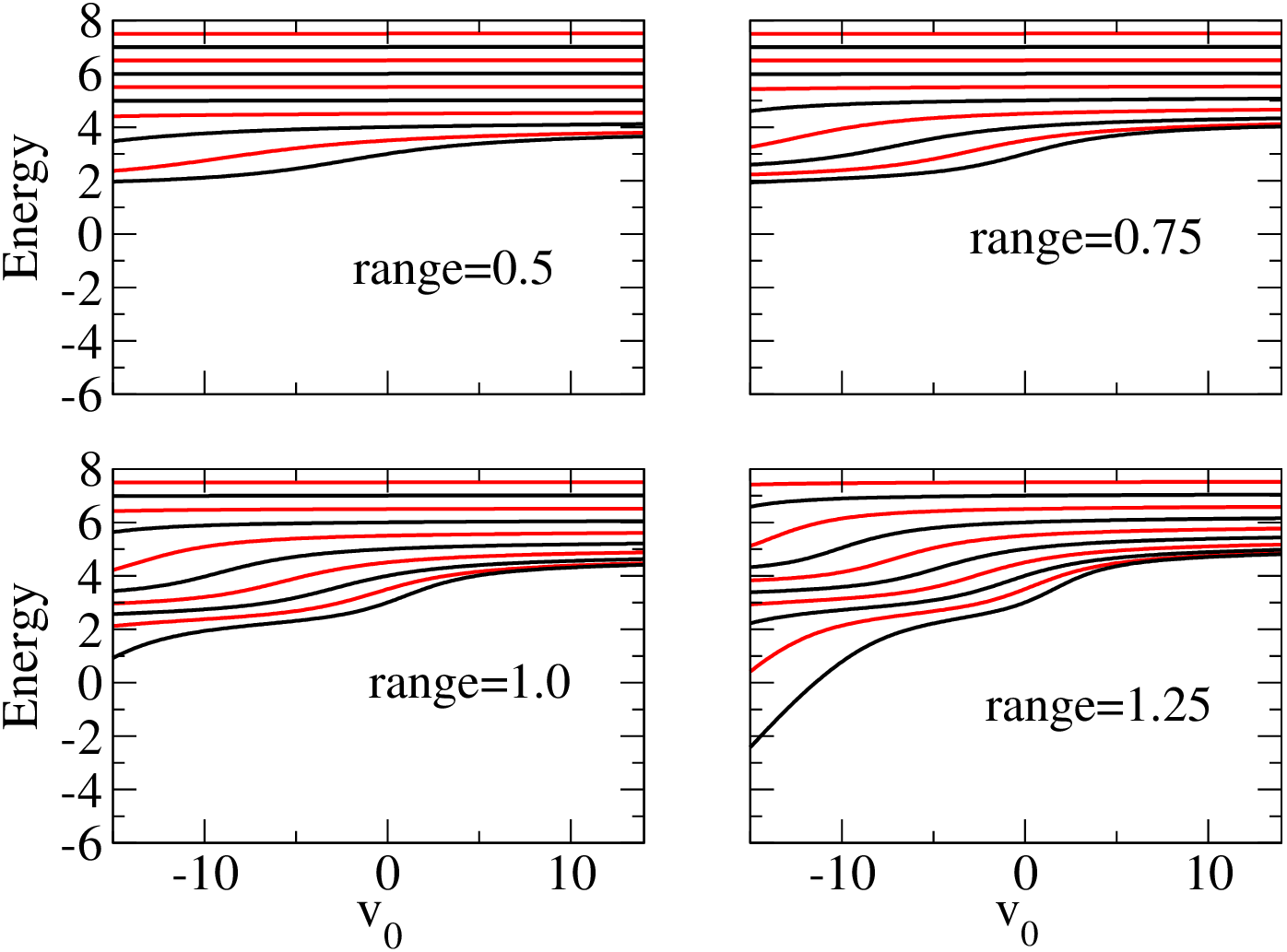}
			\caption{}
			\label{fig:sub20}
		\end{subfigure}
		\caption{\small(a) Energy spectrum for the fundamental state (n=0)  versus $v_0$ with increasing value of the angular momentum quantum number ($l=0$,1,2,3,4 from the bottom to the top) in three (red) and two (black) dimensions . (b) The same calculations for the first excited state $n$=1}
		\label{fig:fig2}
	\end{figure}

	In order to study the effect of the interaction  strength and range, we have elaborated  programs in C language for the three dimensionalities. The continuity equations are resolved using the shooting numerical method \cite{pres,beu} in order to find the eigen energy for different couples ($v_0,a$). The wave function is then used to calculate the probability density and other related quantities.  We follow the prescription given in equation (\ref{eq:eq12}) to extend the results already known for one and two dimensions to three dimensions. Let us now show some of the results we can achieve by exploiting the solutions provided by the previous model, bearing in mind that the comparison is  made relatively to $ v_0=0$  where, we retrieve the simple equidistant spectrum for a harmonic oscillator. It is also important to remind that all the quantities displayed in the figures of the paper are in dimensionless units.  First we can show the effect of the different values of the couple ($v_0,a$) on the energy spectrum for 2D and 3D cases. On figure \ref{fig:fig2}.a, we are depicting the energy of the fundamental level $ n=0$ with different values of the angular momentum quantum number $l$=0,1,2,3 and 4. The energy is plotted versus $v_0$ and the different panels are for different ranges of the interaction. We can see  on this figure that levels with increasing value of angular quantum number $l$ are affected by the interaction as its range is increased. The most important impact is observed when the interaction is attractive (negative $v_0$). In this case the eigen energy is as negative as the interaction is attractive forcing the system to be in a bound state. When redoing the same figure for $n=1$ ( Figure \ref{fig:fig2}.b), one can see that the energy levels are less affected by the interaction and we have to reach a range as high as 1.25 to obtain a noticeable change  for the attractive part of the interaction. It is remarkable however that even in the extreme strength and range of the attractive regime, the increase of the curve is not straight as it is for $n=0$, but proceeds via an alternation of increase and  kind of plateau.  When comparing the results obtained for two dimensions and three dimensions on the same figure, we can notice the regularity with which the levels react to the interaction, whether in the fundamental principal states ($n=0$) on figure \ref{fig:fig2}.a  or the first principal excited states ($n=1$) on figure \ref{fig:fig2}.b. The three dimensions is always higher in energy as the centrifugal potential is naturally higher in this case ($l_{2D}=l_{3D}+1/2$). These results show the effect of the centrifugal potential which scales as the  square of the angular momentum quantum number and acts in a way to repel the system to a separation where it does not feel the effect of the interaction.   
	It is interesting to notice the decrease of the gap between the first curve $l=0$ and the the curve for $l=1$ in the extreme repulsion for  important ranges in two dimensions, making these two level tending towards being degenerate. This result shows that in this regime the repulsion is nearly equal to the amount of the centrifugal potential equivalent to $ l=1$ corresponding to the first fermionic level. This forced degeneracy can lead to the fermionization of the first bosonic state as it is the case for 1D.

	\begin{figure}[h!]
		\centering
		\begin{subfigure}[t]{.5\textwidth}
			\centering
			\includegraphics*[width=.8\linewidth]{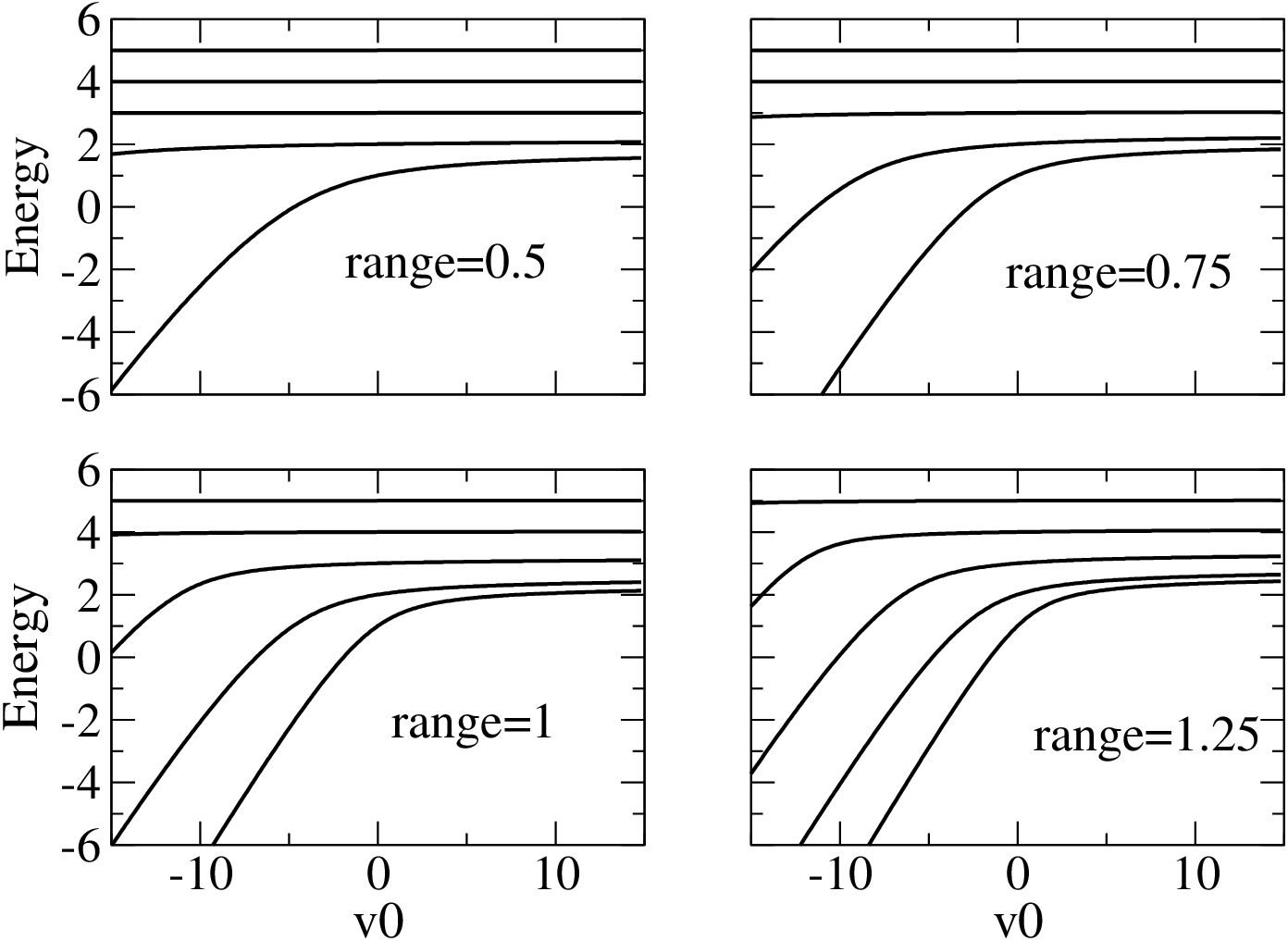}
			\caption{}
			\label{fig:sub11}
		\end{subfigure}%
		\begin{subfigure}[t]{.5\textwidth}
			\centering
			\includegraphics*[width=.8\linewidth]{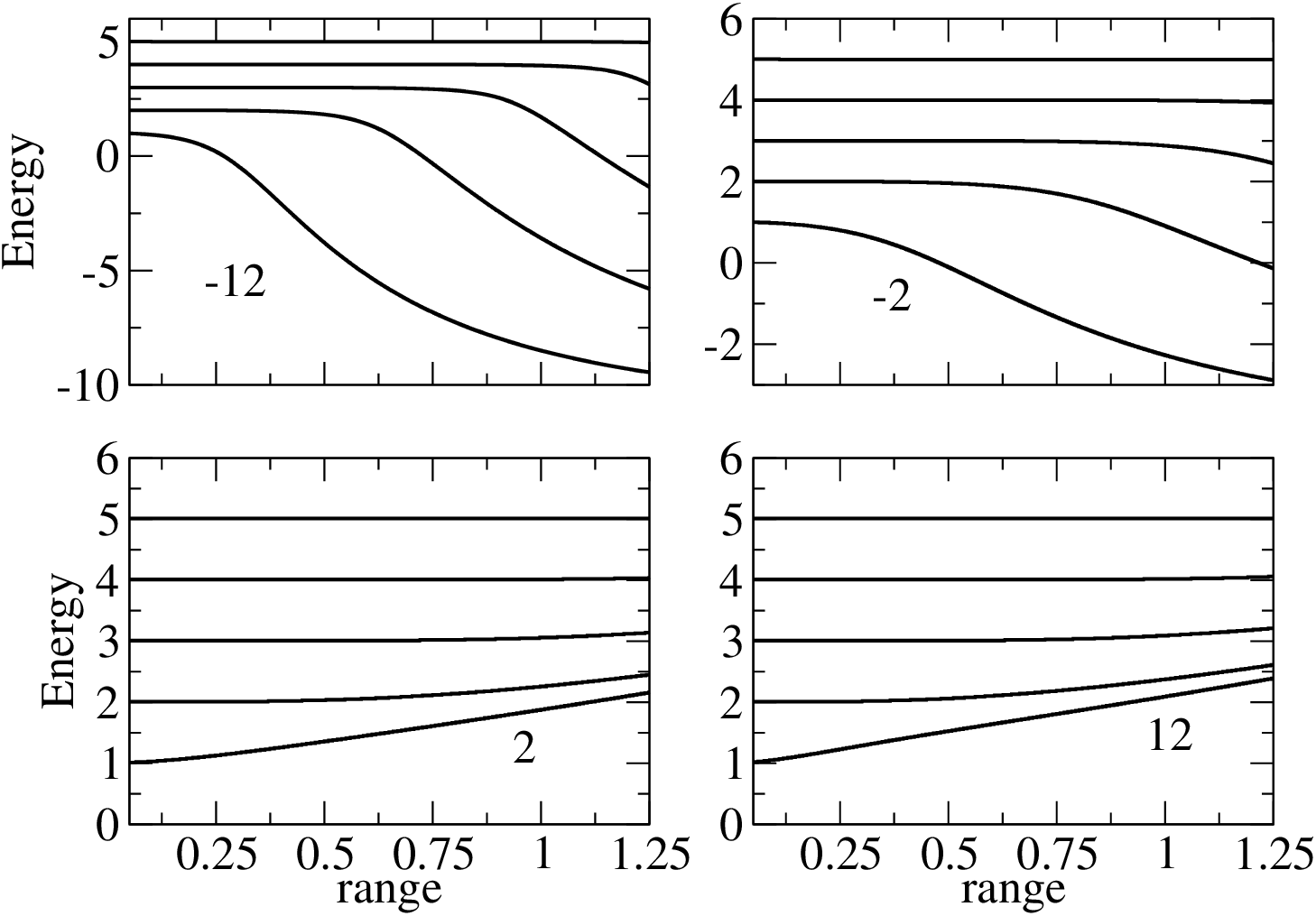}
			\caption{}
			\label{fig:sub21}
		\end{subfigure}
		\caption{(a) Energy spectrum for the fundamental relative-radial state ($n$=0) versus $v_0$ with increasing value of the angular momentum quantum number ($l$=0,1,2,3,4 from the bottom to the top) in two dimensions. The even values of $l$ are for bosons and odd ones are for fermions. (b) The same calculations versus the range for different values of the potential strength . The value of $v_0$ is indicated in each panel}
		\label{fig:fig3}
	\end{figure}

	\begin{figure}[h!]
		\centering
		\begin{subfigure}[t]{0.5\textwidth}
			\centering
			\includegraphics*[width=.8\linewidth]{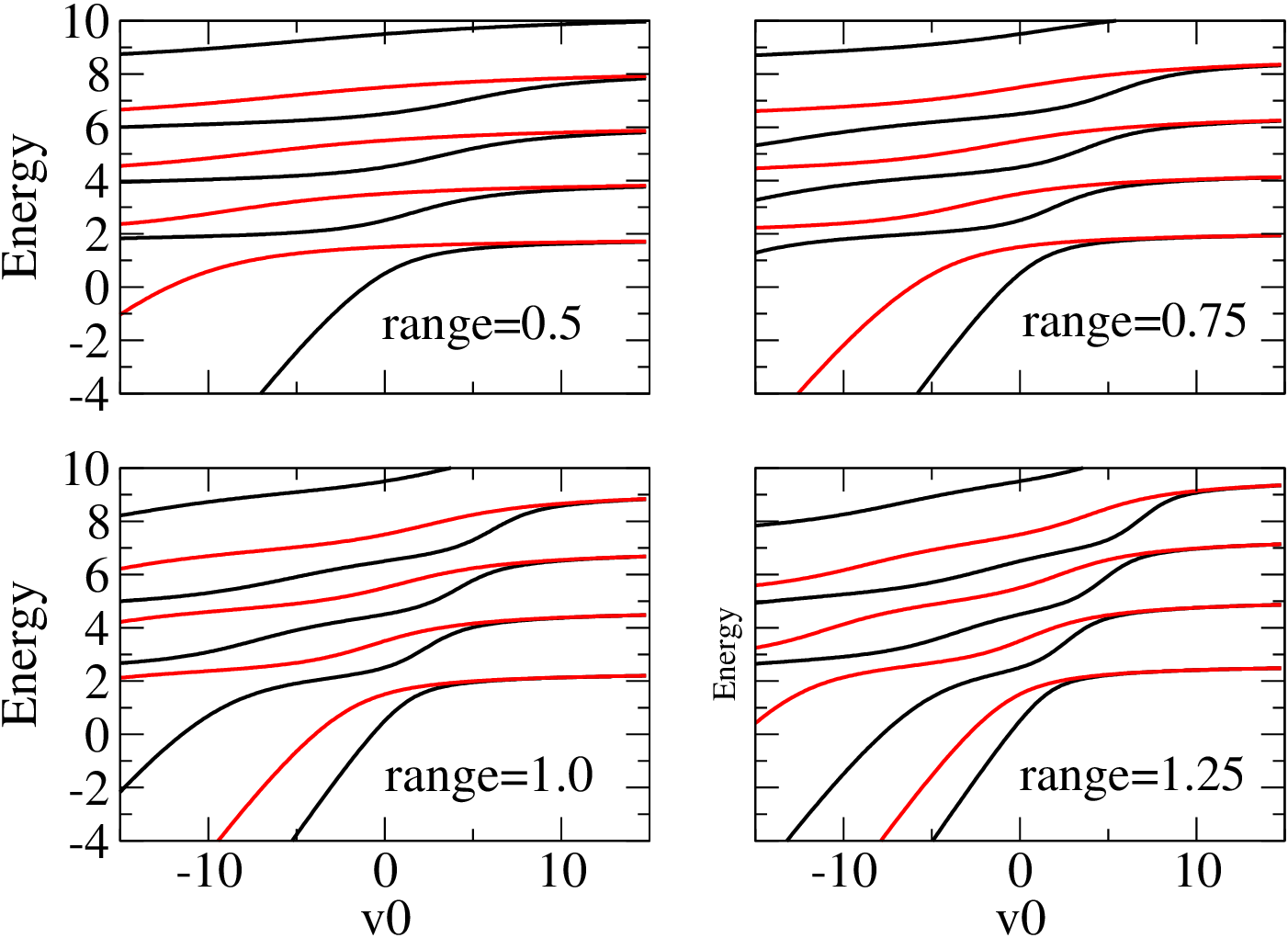}
			\caption{}
			\label{fig:sub12}
		\end{subfigure}%
		\begin{subfigure}[t]{.5\textwidth}
			\centering
			\includegraphics*[width=.8\linewidth]{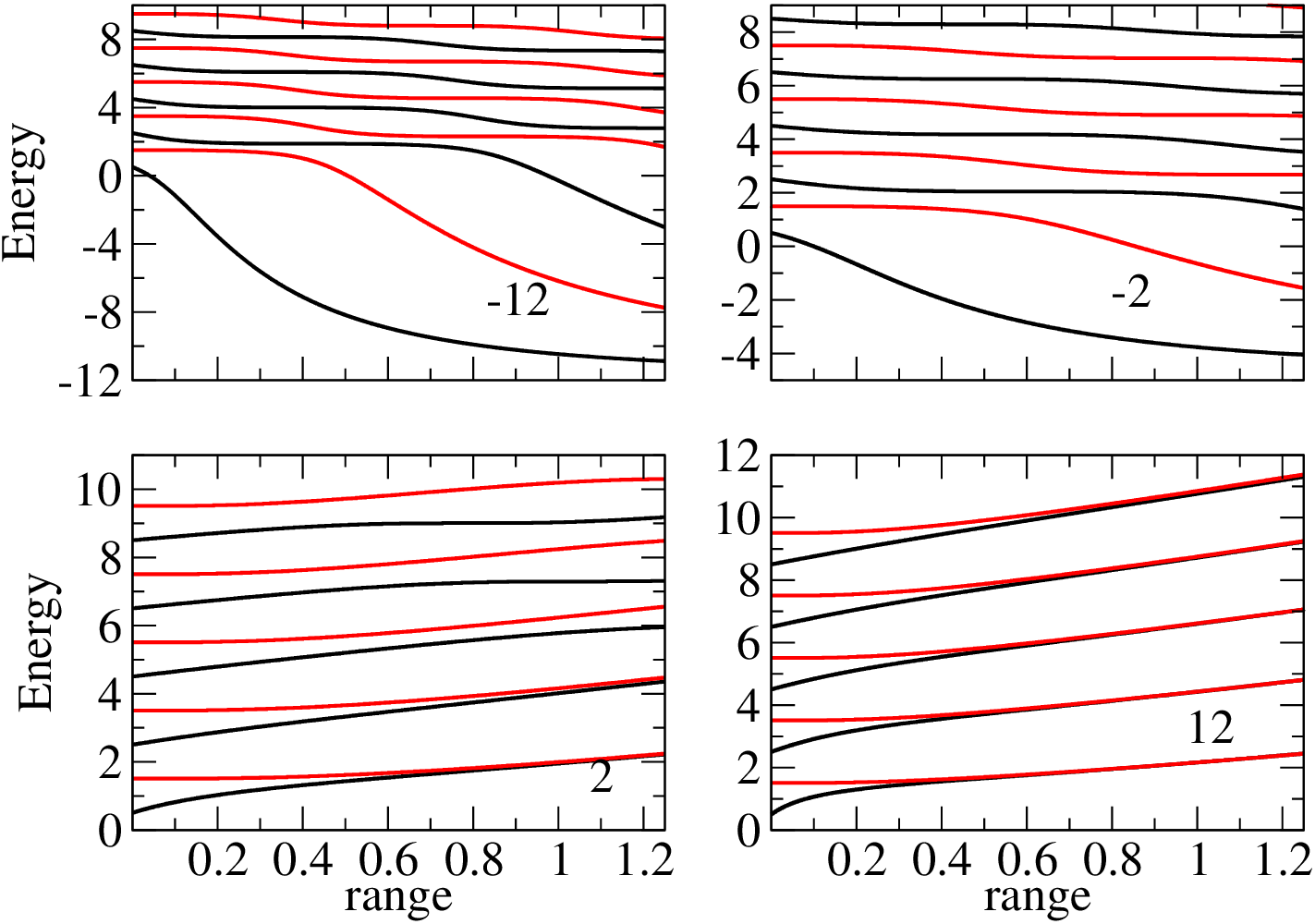}
			\caption{}
			\label{fig:sub22}
		\end{subfigure}
		\caption{(a) Energy spectrum in one dimension versus $v_0$ with increasing value of $n$ ($n$=0,1...8 from the bottom to the top) with even values of $n$ for bosons (black) and odd ones for fermions (red). (b) The same calculation versus the range for different values of the potential strength. The value of $v_0$ is indicated in each panel}
		\label{fig:fig4}
	\end{figure} 
	
	\noindent
	The representation of the energy versus $v_0$ for different values of the range is frequently used to show the effect of the interaction on the energy spectrum. We want to show  in the following figures, the alternate representation: energy spectrum versus the range for different values of  $v_0$. We are comparing the two representations on figure \ref{fig:fig3} for the case of two dimensions. The advantage of representing the energy versus the range over the usually used representation, is that it can show the critical range at which we can observe the onset of any change on the different curves. On figure \ref{fig:fig3}.b and for the attractive regime one can see that the point of inflection of the curves for increasing values of $l$ is increasing gradually. One would expect naively that a difference would exist between the behavior of fermionic and bosonic states, since the first ones are naturally exposed to an additional repulsion due to the Pauli exclusion principle. We can see clearly on the figure that this is not the case and that the main parameter that dictates the critical range at which occurs the inflection is the angular quantum number in connection with the strength of the interaction. This is the case even for the first level ($l$=0 )where the onset of the inflection of the curve is not zero but a certain finite value (see sect.\ref {sec:4} for more details). We state here that the angular momentum is {\it washing out} the effect of the statistics. For the repulsive regime however the centrifugal potential and the interaction act in the same direction and this is occurring in very monotonic manner pushing the energy consequently to a higher values.
	
	\noindent
	These results are to be contrasted with the case of 1D on figure \ref{fig:fig4}.a where the energy is represented versus $v_0$.  We can see that the absence of an angular momentum makes that all the curves are more or less equally affected by the interaction. The effect being dependent only on the range and the strength of the interaction;  and that when the interaction is extremely repulsive the bosons and fermions tend to the same limit. This is related to so called fermionization or the Tonks-Girardeau limit \cite{gira} where the bosons proprieties are similar to the ones for non interacting fermions (except for the impulsion distribution). Notice that these results are already reported in \cite{koci3}.  When using the second representation  (Figure \ref{fig:fig4}.b), one can find that the onset of the inflection  is different when comparing bosonic and fermionic states. Indeed for the first bosonic state the inflection starts from zero whereas for the first fermionic state a certain critical range has to be reached in order for the inflection to occur. For the higher bosonic and fermionic levels, we can observe an evolution which is not straight as it is the case for the first two levels and is tightly related to the change of the interaction strength. The understanding of the behavior of the first bosonic and fermionic levels is quite straightforward and is due to the additional repulsion resulting from the fermionic statistics. We can of course notice that the same curves in the first representation are not purely monotonic as it is the case for the first two levels. For the repulsive  regime and in the second representation we can observe   the same tendency to fermionization except that in this case the limits are not flat as it is seen in the first representation but continue to increase monotonically with increasing value of the range.
	
	\section{Spatial correlations}
	\label{sec:4}
\begin{figure}[h!] 
		\centerline{\includegraphics*[height=3.5in,width=3.7in, keepaspectratio]{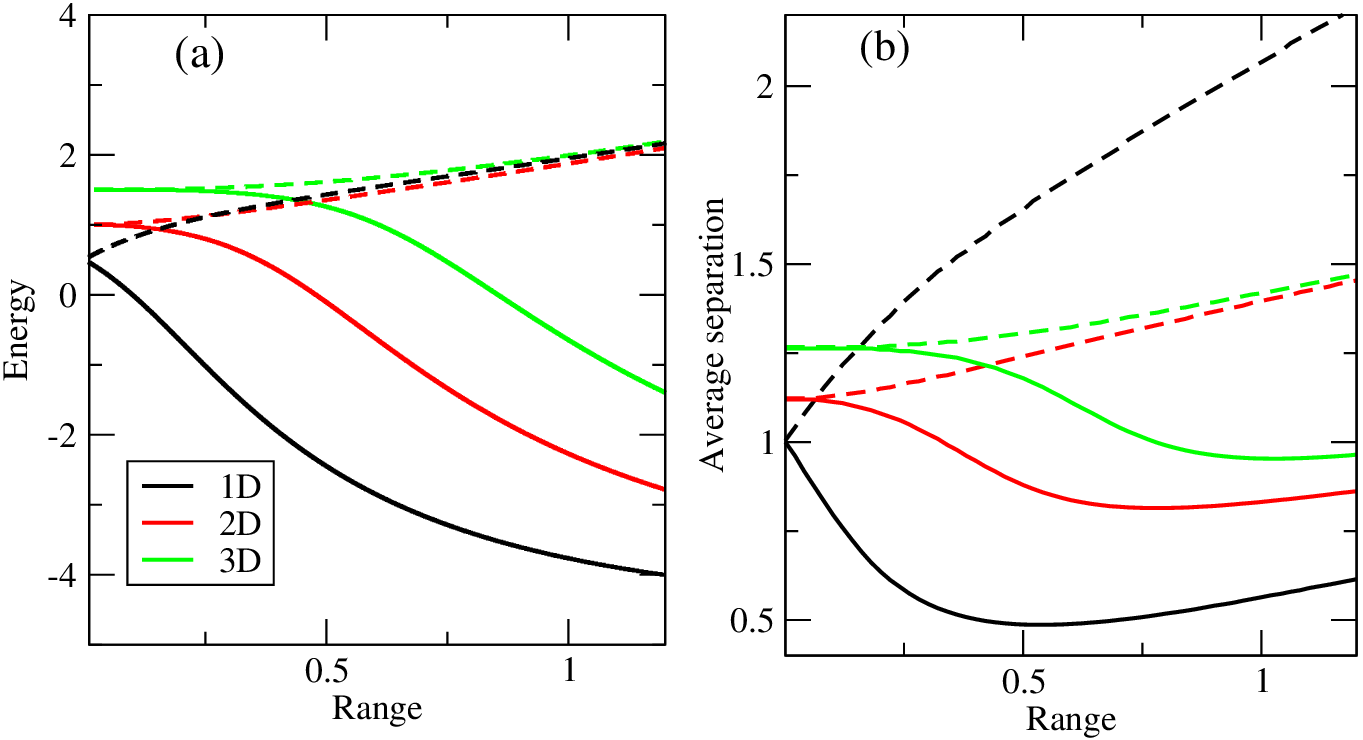} }
		\vspace*{8pt}
		\caption{(a) Comparison of the ground state energy  versus the range of the interaction for the three dimentionalities.(b) comparison of the average separation between the two particles versus the range for the three dimentionalities. The solid curves are for the attractive case ($v_0$=-5) and the dashed curves are for the repulsive case ($v_0$=5).
		}
		\label{fig:fig5}
	\end{figure}

	We intend to show in this section by using the results of the previous model, the impact of the dimentionality on the space correlation of the two particles forming our system. We can see when depicting the energy of the ground state of the relative part of the solution versus the range and for the attractive regime (figure \ref{fig:fig5}.a,  $v_0$=-5, solid lines), that we have a threshold at which the effect of the attraction starts to manifest. We can see clearly that this critical range is different according to the considered dimentionality. The critical range is higher in the 3D case then comes the one for 2D case, whereas the critical range for 1D is equal to zero. The increase in the case of a repulsive regime (figure \ref{fig:fig5}.a $v_0$=5, dashed lines) is more straight with no critical range. The curves in this case are tending peculiarly to the same limit for the three dimentionalities. The different thresholds in the attractive regime is the result of the interplay between the centrifugal repulsion and the attractive interaction. In the case of 1D the centrifugal potential is absent and consequently the critical range is null. For the 3D case even if we set $l=0$ for the ground state, still we are left with  $l=1/2$ in 2D as explained before. Consequently though the centrifugal potential is equal to zero in this case, still $l_{2D}=1/2$ is entering in the arguments of the confluent hypergeometric solution of the 3D equation and consequently this is affecting the solution. Similarly setting $l=0$ in 2D will not annihilate the centrifugal potential since in this case we are left with the residual term $\frac{-1/4}{r^{2}}$. It is clear from the figure \ref{fig:fig5}.a, that the amount of the centrifugal effect is increasing gradually from 1D case to 3D case, passing by the 2D case.

	\noindent
	How these effects are impacting the spatial or pair correlation? To investigate this point, we are plotting the average separation between the two particles versus the range of the interaction for the three dimentionalities (figure \ref{fig:fig5}.b). The average separation is calculated as $\sqrt{\langle x^{2}\rangle}$ and for the average value, we are using the normalized relative-radial part of the wave function. We should mention that the value of the range is assumed for half of the considered space whereas the average separation is calculated on the whole space. This is the case in order for the calculations to be consistent with the previous section where the range for the interaction signify the distance from the origin to the  edge of the step (figure \ref{fig:fig1}).  For the attractive regime and for 1D (black solid curve on  figure \ref{fig:fig5}.b ), the average separation for a range that is nearly null is 1. We are finding the same value of the average separation as when using the wave function of the relative part of the harmonic oscillator without the interaction. For increasing value of the range and for the same case, the average separation is reaching a minimum and then starts to increase. We have to notice here that the decrease occurs straightly without a critical range. The same thing is happening for the cases  2D and 3D (red and green solid curves on figure \ref{fig:fig5}.b ), except that the decrease is occurring   after a certain critical range and the minimum of the two curves is pushed to higher ranges. The critical range and the curve minimum are more important for the 3D case than for the 2D one. These results show that for the 1D case, the average separation is starting from a value where the interaction has no effect and we have to increase the range to the point where it is possible for the system to feel the interaction. After this point the increasing range of the interaction is giving rise gradually to a larger interval where the system is likely to be found. The same explanation holds for the 2D and 3D cases, except that for these cases, the centrifugal effect enters into play. This is resulting in a starting average separation which is higher and consequently a larger range is needed in order for the system to feel the effect of the interaction. Once the minimum is overcome, the difference in the average separation between the three dimentionalities stays nearly constant.  On the figure \ref{fig:fig5}.b, the repulsive regime  ($v_0=5$, dashed curves) for the three dimentionalities is also plotted. In this case, the average  separation increases gradually for the whole extent of the range. The result for 1D is greater than the 2D and 3D that tend to the same limit. We find also that the average separation in 1D is almost proportional  to the range of the interaction. This result reinforces also the impact of the centrifugal effect that scales as  $\frac{1}{r^{2}}$. This effect is acting upon the average separation even at relatively important ranges and is giving the trend of the curves in 2D and 3D. For 1D case, such an effect is absent which makes the interaction impose the average separation according to its range. The spatial correlation for the same system is also studied for 1D in Ref. \cite{koci3} but using the two-particle density profile. The results are shown only for the repulsive regime where the inter-particle distance increases gradually with the increase of the range as confirmed by our calculation. Though our method is unable to localize in which side of the trap are the particles nevertheless, it is able to quantify the average separation on the whole extent of the range for the different dimentionalities and for different regimes in a very simple manner. The method of calculation has the advantage of clarifying  in a very instructive way, the interplay between the interaction features and the centrifugal potential and its impact on the spatial correlation.

	\section{ Perturbation treatment}
	\label{sec:5}
	We shall try to establish in this part in which extent the use of a step function as an approximation for the Rydberg interaction is accurate. We are interested in case of discrepancy between these two results by employing the adequate tools in order to ameliorate the initial model. For this aim we are exploiting the perturbation theory \cite{Cohe} to compare :
	\begin{enumerate}
		\item the results for a step function,
		\item the numerical results for the exact formulation of the potential,
		\item and the results of the perturbation treatment of a step-like potential.  
	\end{enumerate}
	We have to clarify here that in the reference  \cite{koci2}, the numerical (Rydberg interaction) and the analytical (step potential) results are plotted on the same figures showing a discrepancy between these two results. A discrepancy that becomes more apparent for important ranges and in two dimensions. Similarly in reference \cite{dob}, an approximate value of the threshold interaction strength can be calculated analytically for the step potential and compared to the numerical results for the Rydberg interaction. This is done in one dimension and a tiny discrepancy is found between the two results. In our calculation we are not only concerned by reporting the discrepancy that does exist but also by bridging the gap between the situations for the cases of 2D and 1D, using the perturbation tool. This calculation is important from two points of view: reaching an agreement between the two results would confirm from one side, the adequacy of the step function to replace the realistic Rydberg potential as it confirms that the missing part is just a perturbation  and the calculations would from another side, establish a more accurate wave functions basis if a description of few particles systems is targeted .

	\noindent
	To start with, the exact potential is written as :
	\begin{equation}
	%\begin{split}
	v(r)=  \frac{g}{1+\left( \frac{r}{R_c}\right)^6 }= v_s(r)-v_s(r)+ \frac{g}{1+\left( \frac{r}{R_c}\right)^6 } =  v_s(r)+v_{pert},\label{this}   
	% \end{split} 
	\end{equation}
	where, $ v_{pert}(r)=  \frac{g}{1+\left( \frac{r}{R_c}\right)^6 }-v_s(r) $ and $v_s(r)$ is the step function defined in equation (\ref{eq:eq16}).
	This way, it is possible to write the exact potential as a step function for which we already know the  solutions and an an extra quantity $v_{pert}$ that we assume to be a perturbation. A plot for this potential for the case where the step is equal to 5 and a range of the potential is equal to 1, is illustrated on figure \ref{fig:fig1}. The Numerov approach \cite{gian}  is used to obtain the numerical results for the exact potential (the Rydberg potential). Conversely for our case we are targeting an agreement of our calculations with numerical results since these last ones are the best results we can establish for the realistic potential. Let us mention that we use a forward and inward integration method and we impose the continuity of the wave function and its derivative at the turning points\cite{gian,beu}, to ensure the stability of the Numerov calculation. The perturbation correction is assumed to be of the first order for the eigenvalues. To sum up we are comparing solutions of the step potential and these solutions after a perturbation correction with the numerical results.\\  
	
	\begin{figure}[h!]
		\centering
		\begin{subfigure}[b]{.5\textwidth}
			\centering
			\includegraphics*[width=0.95\linewidth]{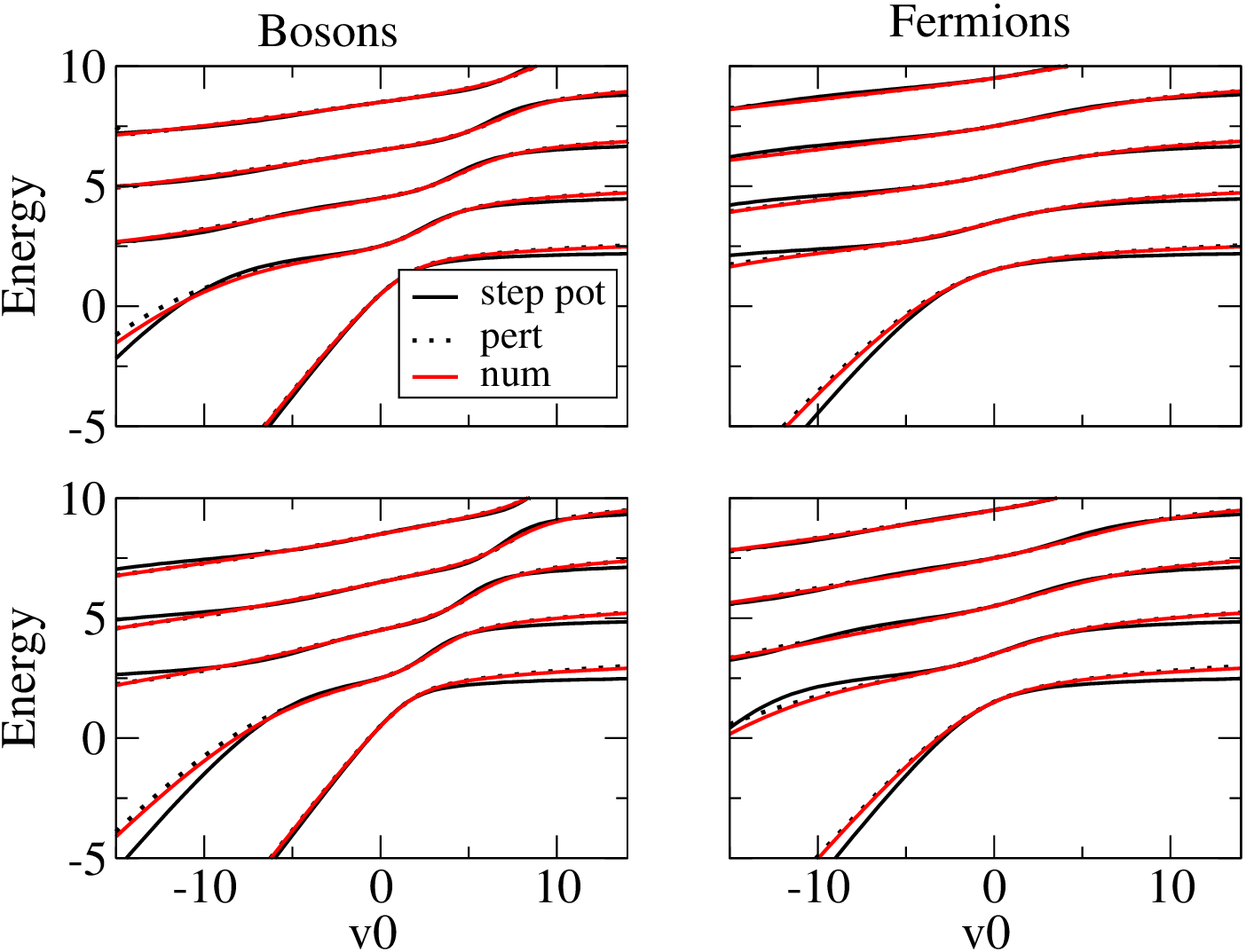}
			\caption{}
			\label{fig:sub1}
		\end{subfigure}%
		\begin{subfigure}[b]{.45\textwidth}
			\centering
			\includegraphics*[width=.5\linewidth]{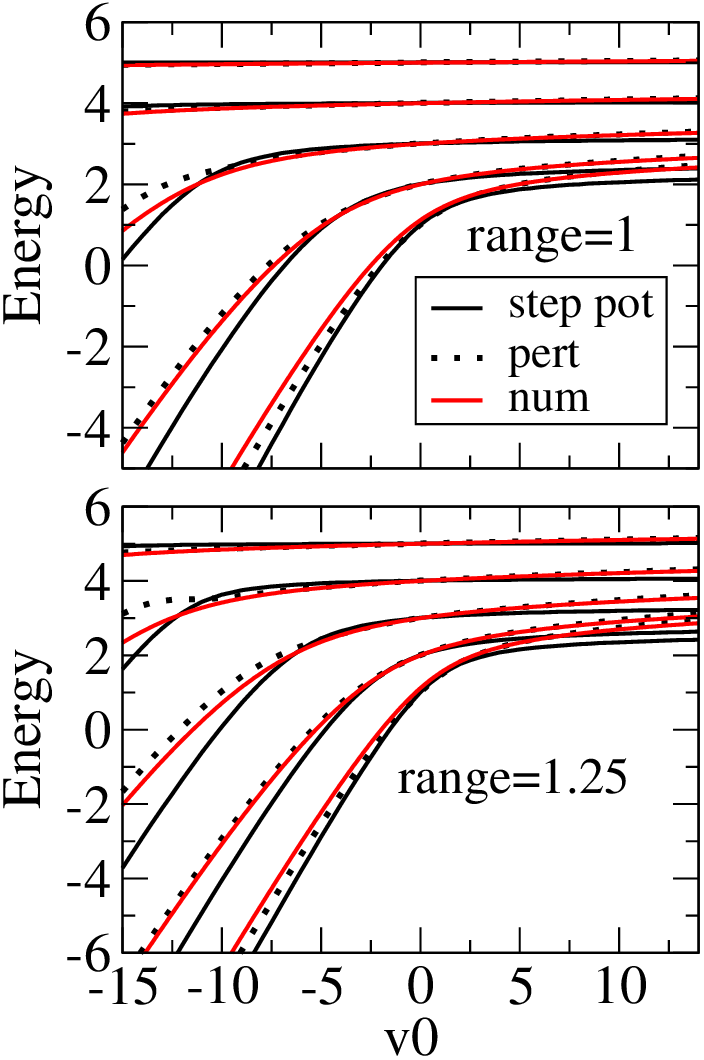}
			\caption{}
			\label{fig:sub2}
		\end{subfigure}
		\caption{ (a) Comparison of the energy versus $ v_0$ for bosons and fermions (columns) for two ranges a=1 and  a=1.25 (rows from the top to the bottom respectively) in 1 dimension. In each panel the calculation for the step function (step pot), perturbation correction (pert) and the numerical results(num) are compared. (b) Comparison of the energy versus $ v_0$ in two dimensions for l=0,1, 2,3 and 4  (from the bottom to the top) for two ranges a=1 and  a=1.25 . Even and odd values of $l$ are for bosons and fermions respectively. }
		\label{fig:fig6}
	\end{figure}

	\noindent
	The comparison of the eigenvalues (energies) versus $v_0$ for one dimension and different ranges is given on figure \ref{fig:fig6}.a. It is clear from this figure that the correction to the first order is sufficient to reach a fair agreement with the numerical results. The higher energy levels are less affected by the interaction according to its range and hence, are already too close to the numerical results. Conversely the low levels are more affected by the interaction and the correction for these levels is quite important. This correction demonstrates an energy level for fermionization which is higher than the one without the correction.
	We tried to operate a perturbation first order correction to the eigenvectors without a noticeable success. It is even found for some values of the range and the strength that the results without correction are closer to the numerical ones. The second order perturbation correction does not ameliorate the situation. In summary we can say that the perturbation treatment to the first order is giving a very satisfactory results for the eigenvalue whereas this treatment is less satisfactory for the eigenvectors especially for important values of the range and the strength of the interaction. \\
	
	\noindent
	In the same manner as before, we extended  the perturbation calculation to the radial part of the Schr\"{o}dinger equation for 2D. The perturbation potential is the same as before. The only difference is the centrifugal term making a logarithmic  mapping  and a transformation of the radial solution  necessary for the densification of the points around zero for the wave function and for recovering the Numerov shape of the equation respectively \cite{gian,beu}. The comparison of the spectra  versus $v_0$ and for different ranges $a$ is illustrated on figure \ref{fig:fig6}.b . We can see on this figure that the correction to the first order for both intermediate and large ranges, is making the agreement with the numerical solution more satisfactory especially for the repulsive regime where the curves are indistinguishable. For the attractive regime the corrected results for lower levels are more satisfactory. 
	The results in 3D are expected to be quite similar to one for 2D as the only difference between the two cases, is an addition in orbital momentum quantum number making the whole spectra to be translated to higher energy. For the eigenvectors we are expecting as before an inadequacy of the perturbation treatment to recover an acceptable agreement with the numerical results.

	\section{Dynamical aspects}
	
\begin{figure}[h!]
		\centerline{\includegraphics*[height=7.5in,width=4.2in, keepaspectratio]{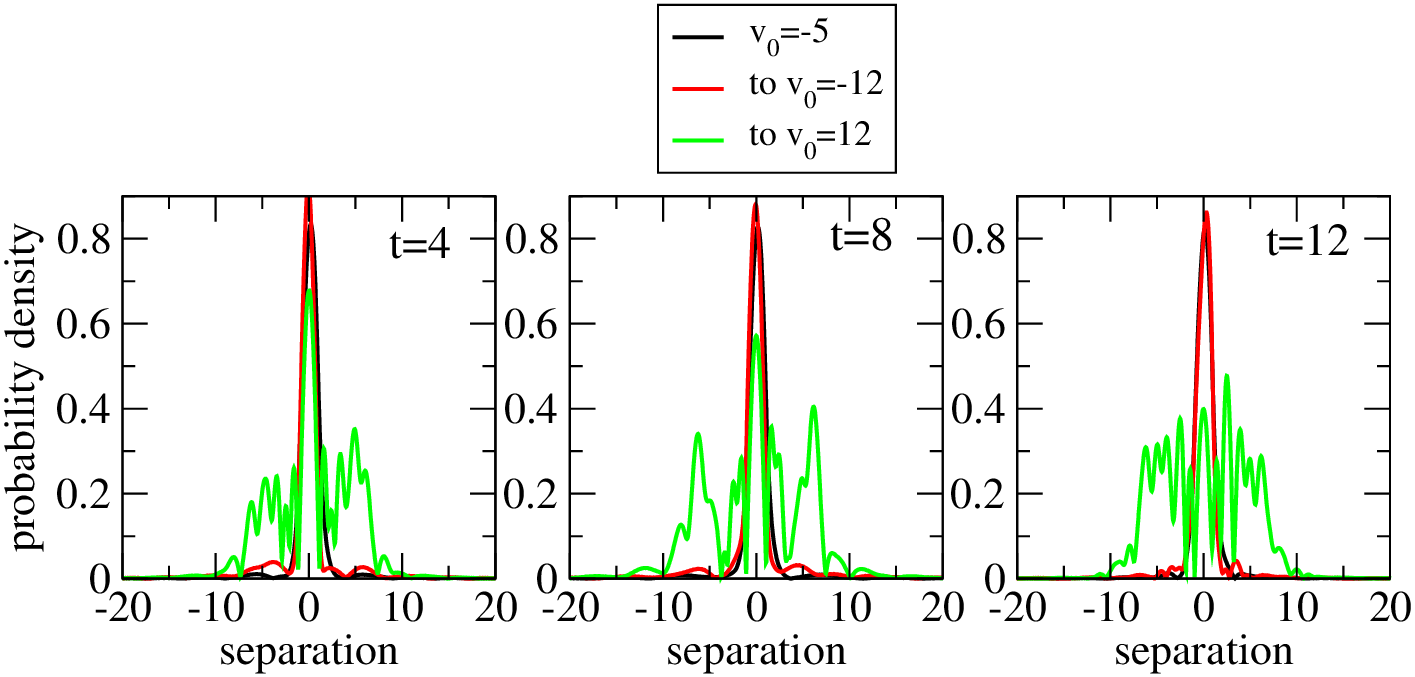} }
		\vspace*{8pt}
		\caption{ Snapshots of the time evolution of the probability density for different scenarios and for one dimension. black line is for a fixed value of $v_0$=-5, red line is for a sudden change of $v_0$ from -5 to -12 and green line is for a sudden change of $v_0$ from -5 to 12. The range is fixed to 1.
		}
		\label{fig:fig7}
	\end{figure}

	A growing interest is presently focusing on studying the evolution of non equilibrium dynamic of cold confined few particles systems. This is leading to a peculiar results that are shedding light on our understanding of the concepts of such problems \cite{dob,keh}. Having at hand analytical solutions for some prototype systems has greatly helped  in investigating some features of the dynamic. Most of these solutions are relying on the contact interaction which is a good candidate in diluted systems \cite{bud}. In the following, we would like to present in a very simplified manner, some of the preliminary results we can achieve by considering our step-like potential for a system of two confined bosons. This interaction choice could be a good candidate in case of strongly correlated systems. The evolution of the system proprieties during time, requires the solution of the time dependent Schr\"{o}dinger equation. We are aiming through this solution to the investigation of the evolution of the system under the initial interaction features compared with the change of the behavior of the system under a sudden change of these same features. This is what is known as quenched interaction. We are elaborating these calculations in one dimension. To solve the time dependent Schr\"{o}dinger equation, we employ the Crank-Nicholson method together with the tridiagonal matrix algorithm exploiting the built-in programs provided by the Lapack library \cite{koo}. We consider grid sizes of $\Delta t$ =0.0002 and $\Delta x$ = 0.04. We take a space of $-30  \leq x \leq $ 30. While the time step is fine enough to avoid any distortion during time, the step and the extent of the space were restrictions of the machine. Nevertheless these are quite satisfactory for the present calculations. The initial wave function from which starts the evolution of the system, is considered to be the exact analytical ground state solution already found by resolving the time independent Shr\"{o}dinger equation for a step potential interaction. We are considering only the relative part of the wave function since the center of motion is not affected by the interaction. This implies that the initial exact ground state solution we have established by resolving the time independent Shr\"{o}dinger equation, is evolved in time, either in the same initial potential or we change suddenly the potential features at t=0. In the first situation  we have just a stationary state and in the second situation, the systems is no longer stationary and it starts to evolve under the new potential.  On figure \ref{fig:fig7} we are illustrating an example of snapshots for given times of the evolution of the probability density. As expected the curves for the stationary  state (black curves in the three panels) stay nearly the same through time. Only very tiny numerical kinks start to develop with time. This is due to the known Crank-Nicolson spurious oscillations that contaminate the wave function at each time iteration without compromising the physical results . When setting a sudden change of the potential strength $v_0$ from -5 to -12 at t=0 (red curves),  the probability is slightly changed but stays located around the first curves. When  in the contrary, the sudden change is operated from -5 to 12, an important oscillation of the density is observed. This is leading to a high non localization of the system. All the previous calculation are elaborated with a fixed value of the interaction range equal to 1.

	\noindent
	It is important to compare results of the contact-like interaction with those for short ranged one. For that, we are comparing the fidelity for two values of the range on the figure \ref{fig:fig7}. The fidelity being the absolute value of the overlap between the initial state of the system and the evolved state. This quantity is a measure of the effect of the quench onto the system and therefore dictates its dynamical response following the quench \cite{bud}. Though the figure \ref{fig:fig7}. a is a representation of the results for a range which is nearly null, it could not be compared directly with known results for a delta-like interaction. This is because we must introduce in  our calculation a scaling of the value of $v_0$ with the strength of the  delta-like potentiel $g_{hc}$ (eq. \ref{eq:eq13}). This is given as : $v_0= 2\times g_{hc}\times range$ \cite{koci3}.  It is not possible to perform such a scaling  with a value of the range that is nearly null and for $ g_{hc}=2$ (a value for which we  already have results \cite{ish} )  without destabilizing the calculation program (arguments beyond the definition domain of the hypergeometric functions). We find it however instructive to present the result for a range that is null without the scaling and compare it with the same calculation for a range that is equal to 1.  We observe that the range is affecting the fidelity in a very pronounced manner. In the case where the range is null, we observe a well defined periodic oscillation between an evolved state that is parallel to the initial state (fidelity =1) and an evolved state that is orthogonal to the initial state (fidelity=0). This behavior could be  compared qualitatively with the results presented in \cite{ish}, though the amplitude of the oscillation is altered for the high strength considered in this reference. In our calculation we can see that the value of $v_0$ is dictating the period of these oscillations. In the case where the range is equal to 1, the behaviour of the curves is completely different. The oscillation of the fidelity becomes irregular between zero and intermediate value between zero and one. Only the sudden change from $v_0$=-5 to 0 is resulting in a periodic oscillation between zero and one.  
	
	\begin{figure}[h!] 
		\centerline{\includegraphics*[height=3.5in,width=3.7in, keepaspectratio]{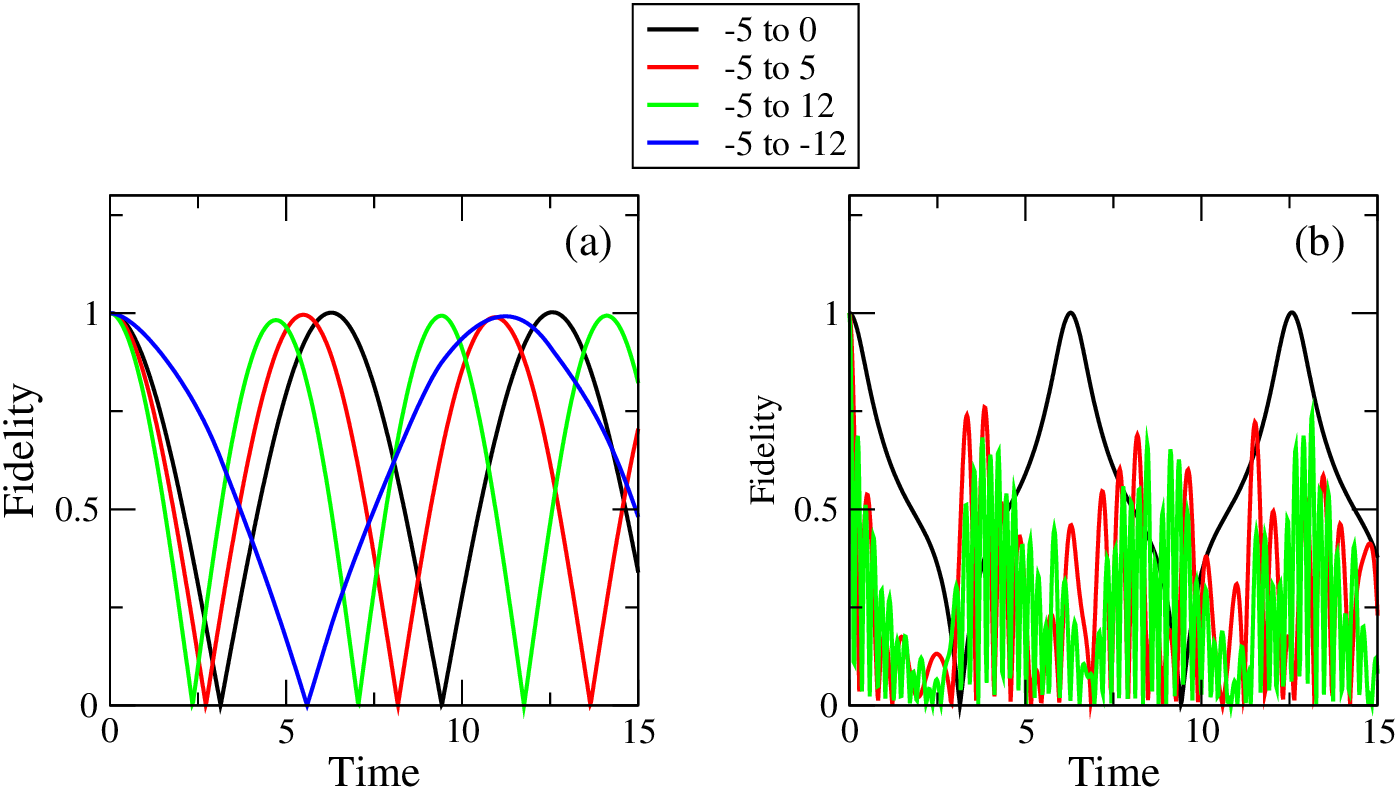} }
		\vspace*{8pt}
		\caption{(a) Evolution of the Fidelity versus time for a range of nearly zero and for different scenarios of the quench of the interaction strength indicated in the legend. (b) The same calculations for a range of 1. The calculation $v_0$ from -5 to -12 is not shown in part (b).
		}
		\label{fig:fig8}
	\end{figure}

	\noindent
	Let us investigate in some details, in which extent the sudden change of the interaction potential is affecting the spatial correlation of the studied system. For that, we are plotting the evolution of the average separation during time for different cases.  For each panel we have the average separation for the indicated potential (black curves i.e stationary state). The other colored curves are giving different scenarios of sudden changes from the initial potential. In each panel, the other two stationary states are indicated in gray. Let us mention first that for the stationary states, we have as expected within the numerical artifact mentioned before, a nearly constant separation for the three potentials ($v_0$=-5, 0 and 5 ). Setting a sudden change of $v_0 $ from -5 to -12 (figure \ref{fig:fig7}.a), is creating a tiny perturbation of the average separation around the initial position. Notice here  that though the change is very important the consequence on the relative position of the two particles is very moderate. When Setting a sudden change of $v_0 $ from -5 to 0 which means that we are shutting off the interaction, a well defined periodic oscillation  of the average separation around a mean position of 1, is observed. This value of the mean position is one we find for a stationary state with $v_0$=0. The sudden transition from $v_0$=-5 to 5 or 12 is creating an important and irregular oscillation of the average separation. These oscillations are seemingly settling down around a value of the average separation which is higher than the initial one. Notice here however that the localization of the system is better for $v_0$=-5 to 5 scenario than for $v_0$=-5 to 12 one. For the panel (b) of the figure \ref{fig:fig7}, the concerned stationary state is the one for $v_0$=0. In this case the sudden change to an attractive potential ( $v_0$=0 to -5 or -12) is creating a tiny perturbation around the initial position. From another side the sudden transition to a repulsive potentials ($v_0$=0 to 5 or 12 ) is creating as in panel (a), an important and an irregular oscillations that are settling down. For the panel (c) the concerned stationary state is the one for $v_0$=5. The sudden passage to the attractive potentials  is creating a perturbation around the initial average position without being able to change dramatically the average separation. The passage to repulsive potential ($v_0$=0 to 12) is resulting in almost regular oscillation of the average separation around a little bit higher position. Shutting off the interaction ($v_0$=-5 to 0) is creating a well defined periodic oscillation around a smaller value of the average separation. The value  of the average separation in this case, is even smaller than the ones we have for  the transition to the attractive potentials.\\
	\noindent
	
	\begin{figure}[h!]
		\centerline{\includegraphics*[height=2.5in,width=7.7in, keepaspectratio]{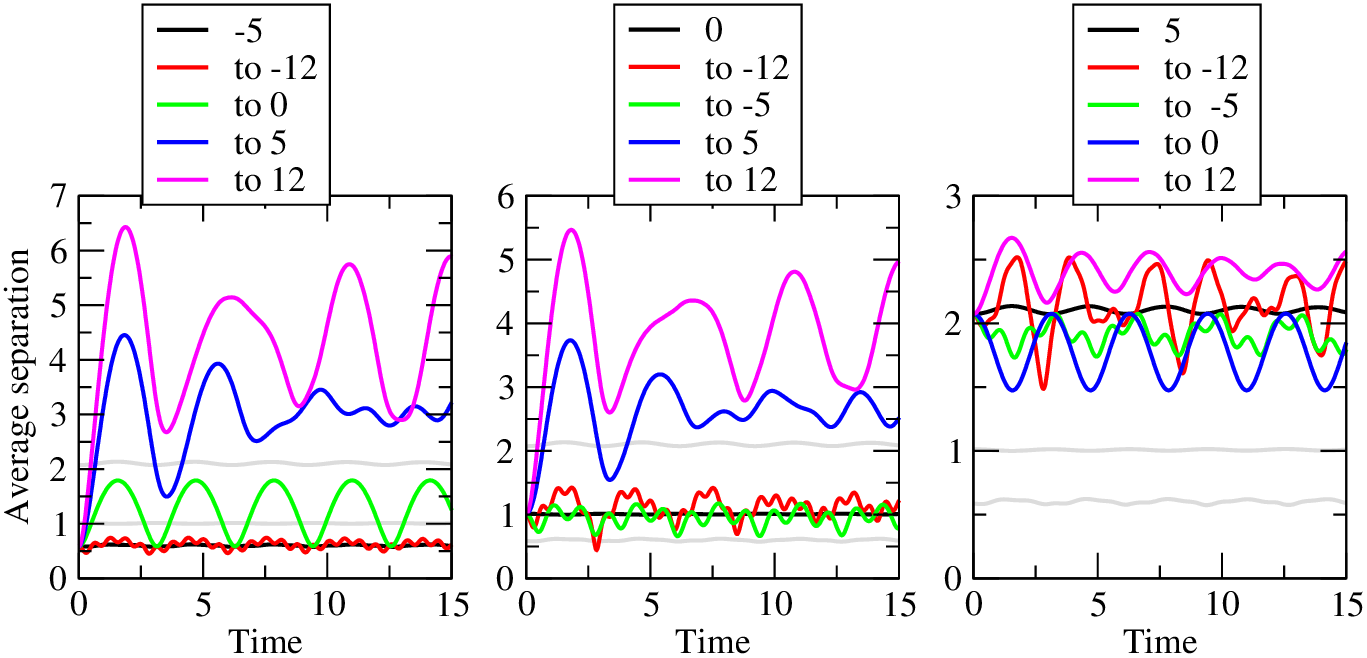} }
		\vspace*{8pt}
		\caption{Evolution of the average separation versus time. The average separation for the stationary states for each fixed potential is indicated in black for the concerned potential and in gray for the two other potentials in each panel. The different scenarios of the sudden change of the potential strength  from the initial value are given in the legend. The range of the interaction is fixed to 1. 
			(a) the initial state is the one for $v_0$= -5, (b) the initial state is the one for $v_0$= 0 and (c) the initial state is one for $v_0$= 5 }
		\label{fig:fig9}
	\end{figure}
	
	\noindent
	To comment on these results, we have to say first that the impact on the stationary state is dependent on the initial state average position as well as the nature of the change of the potential. By the nature of the change, we mean: is it attractive-attractive, attractive-repulsive, repulsive-attractive or repulsive-repulsive transitions? We can notice that when the initial average separation is small (means that the initial state is for an attractive potential) with  an attractive-attractive change, the effect on the average separation is negligible. When for the same average separation, we are setting an attractive-repulsive change, the effect on the average separation is quite noticeable and is related to the magnitude of the final repulsive potential. When on the contrary, the initial average separation is important  (means that the initial state is for a repulsive potential), the passage to an attractive regime does not change greatly the average separation but sets an irregular oscillation around the initial separation. The irregularities are more pronounced for an important magnitude of the final attractive potential. When for the same initial case we are operating a transition to a repulsive regime, nearly the same result is obtained except that in this case, the oscillation is  more regular. It is relatively easy to understand the slight changes for this case, whether for a repulsive-attractive or a repulsive-repulsive transitions, since in this case the mean average separation of the system is nearly out of the reach of the final interaction (range=1 for an average separation of slightly greater than 2).    We have to point out the special case where the transition is leading to the shut off of the potential. The result in this case is a well defined periodic oscillation around a certain mean value.
	
	\section{ Conclusion}
	We presented in this study a detailed comparison of the results concerning the spectra found using the model of Koc{{\' s}}ik. By extending the results to 3D case, it is possible to elaborate a comparative study for the three dimensionalities. It is clearly established that the response of the spectra in 1D case is dependent on the range and the strength of the interaction. Conversely in 2D and 3D cases, the centrifugal potential scaling as the square of the angular quantum number can push the particles to a separation where the effect of the interaction is negligible and consequently the equidistant spectra are conserved. The most important change in the spectra is seen for the attractive part of the interaction. By exploiting the probability density distribution, it is possible to study the spatial correlation for different schemes and dimensionalities. We can say that as expected the spatial correlation for the considered system is tightly dependent on an interplay between a centrifugal effect imposed by the system and a coupled (range, strength) effect imposed by the interaction. The effect is more pronounced in the attractive regime interaction as the interaction and the centrifugal effect in this case are antagonist. This result can be obtained only with finite range interactions and is not be possible with contact or hard core interactions especially for fermionic states where contact interaction is de facto, forbidden . The interesting result is that even in the case of the first bosonic state (the ground state) where the angular momentum quantum number is zero, the difference between the three dimensionalities is shown to manifest as a \enquote{hidden} proper amount of centrifugal effect. An effect that is impacting the results for short and relatively long ranged interaction.  The perturbation treatment of the potential allowed an improvement of the model of Ko{{\' s}}cik concerning the eigenvalues where a satisfactory results are obtained in the three dimensionalities. This is reinforcing the validity of the step function approximation as it demonstrates that the difference between the step function potential and the realistic interaction is just a perturbation that can be recovered by the conventional perturbation treatment. This is not the case for the eigenvectors where unfortunately the results are far from being satisfactory. The study of the dynamical evolution of the system under a sudden change of the potential allowed gaining some insights into the system behavior. It is clear from our preliminary results that the ranged interaction could have an impact completely different from the ones we have for a contact-like interaction. A more deep investigation along this axis could shed more light on some fundamental aspects related to strongly correlated systems as well as eventually provide some experimental clues on how to monitor the system correlation.

	\section*{Acknowledgments}
	One of the authors (N.G) is grateful to N. Rowley for the help he friendly brought for the visit to IPN, Orsay. Many thanks also to D. Lacroix for the warm welcome to IPN, Orsay and for all the help. 
	\bibliographystyle{unsrturl}
	\bibliography{doc} 

\begin{thebibliography}{10}

\bibitem{blum}
D.~Blume{,} \enquote{Few-body physics with ultracold atomic and molecular
  systems in traps}{,} \textit{ Reports on Progress in Physics}~\textbf{75}
  (2012)~046401.

\bibitem{abra}
J.~W. Abraham{,} M.~Bonitz{,} \enquote{Quantum Breathing Mode of Trapped
  Particles: From Nanoplasmas to Ultracold Gases}{,}
  \href{https://doi.org/10.1002/ctpp.201300066}{\textit {contributions to
  plasma physics} \textbf{54} (2014) 27 }.

\bibitem{turb}
A.~V.~Turbiner{,} \enquote{One-dimensional quasi-exactly solvable {S}chrodinger
  equations}{,} \textit{ Physics Reports}~\textbf{642} (2016) 1-71.

\bibitem{mora}
J.~Morales{,} J. Garcia-Martínez{,} J. Garcia-Ravelo{,} J. J.~Pen{\~a}{,}
  \enquote{Exactly solvable {S}chrodinger equation with hypergeometric
  wavefunctions}{,} \textit{Journal of Applied Mathematics and
  Physics}~\textbf{3} (2015)~454.

\bibitem{gao1}
B.~Gao{,} \enquote{Solutions of the {S}chrodinger equation for an attractive
  1/$r^6$ potential}{,} \textit{ Physical Review A}~\textbf{58} (1998)~1728.

\bibitem{gao2}
B.~Gao{,} \enquote{Repulsive 1/$r^3$ interaction}{,} \textit{ Physical Review
  A}~\textbf{ 59} (1999)~2778.

\bibitem{gino}
J.N.~Ginocchio{,} \enquote{A class of exactly solvable potentials{,}
  One-dimensional {S}chrodinger equation}{,}
  \href{https://doi.org/10.1016/0003-4916(84)90084-8}{\textit {Annals of
  Physics} \textbf{152} (1984) 203-219 }.

\bibitem{blai}
J-P. Blaizot{,} G.~Ripka{,} \enquote{Quantum Theory of Finite
  Systems}{,}~massachusett: MIT Press~(1985).

\bibitem{pita}
L.~Pitaevskii{,} S.~Stringari{,} \enquote{Bose-Einstein condensation}{,}
  Oxford: Clarendon Press~(2003).

\bibitem{koci1}
P.~Ko{{\' s}}cik{,} A. Kuro{{\' s}}{,} A. Pieprzycki{,} T.~Sowi{{\' n}}ski{,}
  \enquote{Pair correlation ansatz for the ground state of interacting bosons
  in an arbitrary one dimensional potential}{,}
  \href{https://doi.org/10.1038/s41598-021-92556-7}{\textit {scientific
  reports} \textbf{11} (2021) 13168 }.

\bibitem{busc}
T.~Busch{,} B-G. Englert{,} K. Rzazewski{,} M.~Wilkens{,} \enquote{Two cold
  atoms in a harmonic trap}{,} \textit{Foundations of Physics}~\textbf{ 28}
  (1998)~549–559.

\bibitem{wei}
B-B.~Wei{,} \enquote{Two one-dimensional interacting particles in a harmonic
  trap}{,} \href{https://doi.org/10.1142/S0217979209053345}{\textit
  {International Journal of Modern Physics B} \textbf{23} (2009) 3709 }.

\bibitem{doga}
R.~A. Doganov{,} S. Klaiman{,} O. E. Alon{,} A. I. Streltsov{,} L
  S~Cederbaum{,} \enquote{Two trapped particles interacting by a finite-range
  two-body potential in two spatial dimensions}{,}
  \href{https://doi.org/10.1103/PhysRevA.87.033631}{\textit {Physical Review A}
  \textbf{87} (2013) 033631 }.

\bibitem{muja}
P.~Mujal{,} A. Polls{,} B. Juliá-Díaz{,} \enquote{Fermionic properties of two
  interacting bosons in a two-dimensional harmonic trap}{,}
  \href{https://doi.org/10.3390/condmat3010009}{\textit {Condensed Matter}
  \textbf{3} (2018) 9 }.

\bibitem{oldz}
R.~O{{\l}}dziejewski{,} W. G{{\' o}}recki{,} K.~Rza{{\' z}}ewski{,}
  \enquote{Two dipolar atoms in a harmonic trap}{,}
  \href{https://doi.org/10.1209/0295-5075/114/46003}{\textit{Europhysics
  Letters} \textbf{114} (2016) 4 }.

\bibitem{lim}
J.~Lim{,} H-G. Lee{,} J.~Ahn{,} \enquote{Review of cold Rydberg atoms and their
  applications}{,} \textit{ Journal of the Korean Physical Society}~\textbf{
  63} (2013)~867–876.

\bibitem{koci2}
P.~Ko{{\' s}}cik{,} T.~Sowi{{\' n}}ski{,} \enquote{Exactly solvable model of
  two interacting Rydberg-dressed atoms confined in a two-dimensional harmonic
  trap}{,} \href{https://doi.org/10.1038/s41598-019-48442-4}{\textit
  {scientific reports} \textbf{8} (2019) 12018 }.

\bibitem{zinn}
N.~T.~Zinner{,} \enquote{Exploring the few- to many-body crossover using cold
  atoms in one dimension}{,}
  \href{https://doi.org/10.1051/epjconf/201611301002}{\textit {EPJ Web of
  Conferences} \textbf{113} (2016) 01002 }.

\bibitem{Isla}
R.~Islam{,} C. Senko{,} W. C. Campbell{,} S. Korenblit{,} J.~Smith
  et~al.{,}~\enquote{Emergence and Frustration of magnetism with variable-range
  interactions in a quantum simulator}{,}
  \href{10.1126/science.1232296}{\textit {Science} \textbf{340} (2013) 583-587
  }.

\bibitem{gor}
A.~G\"{o}rlitz{,} J. M. Vogels{,} A. E. Leanhardt{,} C. Raman{,} T.
  L.~Gustavson et~al.{,}~\enquote{Realization of Bose-Einstein condensates in
  lower dimensions}{,}
  \href{https://doi.org/10.1103/PhysRevLett.87.130402}{\textit {Physical Review
  Letter} \textbf{87} (2001) 130402}.

\bibitem{dali}
J.~Dalibard{,} \enquote{Les interactions entre atomes dans les gaz
  quantiques}{,} Collège de France: lecture notes (2020)~(in French).

\bibitem{math}
Jr. W. N. Mathews{,} M. A. Esrick{,} Z. Teoh{,} J. K. Freericks~{,} \enquote{A
  physicist's guide to the solution of Kummer's equation and confluent
  hypergeometric functions}{,}~\href{
  https://doi.org/10.48550/arXiv.2111.04852} { arXiv:2111.04852 (2021)}.

\bibitem{abra2}
M.~Abramowitz{,} I.~Stegan{,} \enquote{Handbook of mathematical functions with
  formulas{,} graphs{,} and mathematical tables}{,} Washington: U.S. Government
  Printing Office~(1964).

\bibitem{digi}
\enquote{Digital Library of Mathematical Functions}{,}
  \href{http://dlmf.nist.gov/}{ DLMF} (2021).

\bibitem{pres}
W.~H. Press{,} S. A. Teukolsky{,} W. T. Vetterling{,} B. P.~Flannery{,}
  \enquote{Numerical recipes{,} the art of scientific computing}{,} Cambridge:
  Cambridge Press~(2007).

\bibitem{beu}
T.~A.~Beu{,} \enquote{Introduction to numerical programming{,} a practical
  guide for scientists and engineers using Python and C/C++}{,} Florida: CRC
  Press~(2015).

\bibitem{gira}
M.~Girardeau{,} \enquote{Relationship between systems of impenetrable bosons
  and fermions in one dimension}{,}
  \href{https://doi.org/10.1063/1.1703687}{\textit {Journal of Mathematical
  Physics} \textbf{1} (1960) 516 }.

\bibitem{koci3}
P.~Ko{{\' s}}cik{,} T.~Sowi{{\' n}}ski{,} \enquote{Exactly solvable model of
  two trapped quantum particles interacting via finite-range soft-core
  interactions}{,} \href{https://doi.org/10.1038/s41598-017-18505-5}{\textit
  {Scientific report} \textbf{8} (2018) 48 }.

\bibitem{Cohe}
C.~Cohen-Tannoudji{,} B. Dui{,} F.~Laloe{,} \enquote{Quantum mechanics}{,}
  Cambridge: Wiley~(1991).

\bibitem{dob}
J~Dobrzyniecki{,} T~Sowi{{\' n}}ski{,} \enquote{Two Rydberg-dressed atoms
  escaping from an open well}{,}
  \href{https://doi.org/10.1103/PhysRevA.103.013304}{\textit {Physical Review
  A} \textbf{103} (2021) 013304 }.

\bibitem{gian}
P.~Giannozzi{,} F. Ercolessi{,} S. D.~Gironcoli{,} \enquote{Numerical methods
  in quantum mechanic}{,} Trieste: lecture~notes (2021).

\bibitem{keh}
L.~M. A. Kehrberger{,} V. J. Bolsinger{,} P.~Schmelcher{,} \enquote{Quantum
  dynamics of two trapped bosons following infinite interaction quenches}{,}
  \href{https://doi.org/10.1103/PhysRevA.97.013606}{\textit {Physical Review A}
  \textbf{97} (2018) 013606}.

\bibitem{bud}
L.~Budewig{,} S. I. Mistakidis{,} P.~Schmelcher{,} \enquote{Quench dynamics of
  two one-dimensional harmonically trapped bosons bridging attraction and
  repulsion}{,} \href{https://doi.org/10.1080/00268976.2019.1575995}{\textit
  {Molecular Physics} \textbf{117} (2019) 2043-2057}.

\bibitem{koo}
S.E. Koonin{,} D.C.~Meredith{,} \enquote{Computational physics: Fortran
  Version}{,} Cambridge: Addison-Wesley Publishing Company~(1990).

\bibitem{ish}
I.~S.~Ishmukhamedov{,} \enquote{Quench dynamics of two interacting atoms in a
  one-dimensional anharmonic trap}{,}
  \href{https://doi.org/10.1016/j.physe.2022.115228}{\textit {Physica E:
  Low-dimensional Systems and Nanostructures} \textbf{142} (2022) 115228}.

\end{thebibliography}
	
\end{document}